\documentclass[11pt,twocolumn]{article}

\usepackage{chapterbib}
\usepackage{times}
\usepackage{geometry}
\usepackage{graphicx}
\usepackage{lineno}
\usepackage{amsmath,amssymb}
\usepackage{subcaption}
\usepackage{xcolor}
\newcommand{\rev}[1]{\textcolor{black}{#1}}

\newcommand{\bra}[1]{\langle#1 |}
\newcommand{\ket}[1]{|#1 \rangle}

\newcommand{\innerprod}[2]{\left \langle #1 | #2 \right\rangle}

\begin{document}

\title{Integrating Quantum Algorithms Into Classical Frameworks: A Predictor-corrector Approach Using HHL}
\author{
 Omer Rathore$^{1*}$, Alastair Basden$^{1}$, Nicholas Chancellor$^{1,2}$, Halim Kusumaatmaja$^{1,3}$ \\
$^{1}$ Department of Physics, Durham University, Durham DH1 3LE, United Kingdom \\
 $^{2}$ School of Computing, Newcastle University, Newcastle upon Tyne NE4 5TG, United Kingdom \\
  $^{3}$ Institute for Multiscale Thermofluids, School of Engineering, University of Edinburgh,\\
  Edinburgh EH9 3FD, United Kingdom \\
 omer.rathore@durham.ac.uk \\ 
 a.g.basden@durham.ac.uk, nicholas.chancellor@gmail.com, halim.kusumaatmaja@ed.ac.uk \\
}
\date{}
\twocolumn[
  \begin{@twocolumnfalse}
    \vspace{-\baselineskip}
    \vspace{-\baselineskip}
    \maketitle
   \vspace{-\baselineskip}
   \vspace{-\baselineskip}
   
    \begin{abstract}
The application of quantum algorithms to classical problems is generally accompanied by significant bottlenecks when transferring data between quantum and classical states, often negating any intrinsic quantum advantage.  
\rev{Here we address this challenge for a well-known algorithm for linear systems of equations, originally proposed by Harrow, Hassidim and Lloyd (HHL), by adapting it into a predictor-corrector instead of a direct solver.}
Rather than seeking the solution at the next time step, the goal now becomes determining the change between time steps. This strategy enables the intelligent omission of computationally costly steps commonly found in many classical algorithms, while simultaneously mitigating the notorious readout problems associated with extracting solutions from a quantum state. \rev{Random or regularly performed skips instead lead to simulation failure.} We demonstrate that our methodology secures a useful polynomial advantage over a conventional application of the HHL algorithm. The practicality and versatility of the approach are illustrated through applications in various fields such as smoothed particle hydrodynamics, plasma simulations, and reactive flow configurations. Moreover, the proposed algorithm is well suited to run asynchronously on future heterogeneous hardware infrastructures and can effectively leverage the synergistic strengths of classical as well as quantum compute resources. 
      \vspace{0.5cm} 
    \end{abstract}
  \end{@twocolumnfalse}
]

\section*{Introduction}
The genesis of quantum computing, envisioned over four decades ago as a means to simulate complex quantum problems \cite{feynman1982simulating}, rapidly evolved with the subsequent introduction of a theoretical framework for what such a machine might entail \cite{deutsch1985quantum}. 
This established the foundation for \rev{the field of quantum computing}, driven by the desire to harness quantum mechanical phenomena such as superposition and entanglement, to surpass the capabilities of classical computing. 
Motivated by the potential of a quantum advantage, quantum computing has witnessed continuous growth and has now permeated into numerous fields that include biology \cite{outeiral2021prospects}, weather forecasting \cite{tennie2023quantum}, engineering \cite{national2019quantum}, finance \cite{orus2019quantum} and drug discovery \cite{cao2018potential}. 

Recent advances in quantum hardware \cite{tacchino2020quantum,de2021materials} have further complemented an already burgeoning interest in the field, fueled by seminal quantum algorithms, which best showcase quantum supremacy over classical approaches.
Among these, Grover's quantum search method \cite{grover1996fast} and Shor's algorithm for factoring large numbers \cite{shor1999polynomial} stand out as pivotal milestones. Fundamentally, these algorithms rely on quantum phase estimation \cite{kitaev1995quantum} and amplitude amplification \cite{brassard2002quantum}, which together also form the core of an algorithm to solve linear systems of equations investigated here.      

In 2009, Harrow, Hassidim, and Lloyd proposed a quantum algorithm \cite{harrow2009quantum}, commonly referred to as HHL, that promised to solve systems of linear equations exponentially faster than classical methods. 
The HHL algorithm sparked widespread excitement within the scientific community, given the ubiquitous role of linear equations across numerous disciplines.
For instance, archetypal mathematical operations like matrix inversion, integral to various computational tasks, can be recast as linear systems, making any potential for speedup profoundly useful. 
This potential is embodied by the plethora of applications that employ HHL, across domains such as chemical kinetics \cite{akiba2023carleman}, regression studies \cite{wang2017quantum}, computational fluid dynamics \cite{lapworth2022hybrid} and quantum machine learning \cite{duan2020survey}.  

Numerous studies have sought to improve the HHL algorithm directly, whether by introducing variable time amplitude amplification \cite{ambainis2012variable}, bypassing the quantum phase estimation \cite{childs2017quantum} or improved circuit design \cite{zhang2022improved,yalovetzky2021hybrid}. 
However, this paper adopts a different approach, focusing instead on the broader aspect of how the algorithm is applied. This ensures compatibility with HHL's different variations and also retains sufficient generality to be relevant for other quantum algorithms. 
By critically examining the inherent limitations of the algorithm, this work proposes a paradigm shift: utilizing HHL as a predictor-corrector mechanism.

Our interest in adopting a predictor-corrector approach is motivated by the observation that linear system problems (LSPs) often constitute only a segment of the larger algorithm, yet are resolved at every time step. 
However, in numerous classical scenarios a LSP characterizes a facet of the underlying physics that remains relatively consistent or only varies slowly across time steps. Therefore, it is plausible to suggest that bypassing the resolution of this LSP would not significantly degrade the overall quality of the solution. 

Thus, the pivotal question shifts from determining the solution at the subsequent time step to assessing how much the next solution deviates from the current one. If the solutions between time steps exhibit substantial differences, the LSP is directly solved using classical methods and a full solution \rev{is} obtained. 
Conversely, if the solutions are sufficiently similar, the computationally intensive full classical solve can be circumvented. \rev{In doing so we best leverage the inherently quantum ability of HHL to sample from a population without explicit knowledge of the entire solution, resulting in a novel type of predictor-corrector. Moreover, such a speedup cannot be reliably reproduced by randomly or regularly skipping resolution of the LSP.}

This conceptual shift in how HHL is integrated into a larger algorithm is motivated by the fact that actually harnessing a quantum advantage in practice is very nuanced  \cite{aaronson2015read}, particularly regarding caveats on matrix sparsity, conditioning, as well as the intricacies of data input and output. Without careful application, the potential exponential speedup of quantum computing can be easily squandered, and careful budgeting needs to be employed even for a polynomial advantage. This is in contrast to combinatorial optimisation and quantum chemistry applications, where classical runtimes are expected to scale exponentially, so keeping track of polynomial factors would not be crucial for demonstrating an exponential advantage. Chemistry applications seem like a natural setting for an exponential advantage, given that many of the problems amount to simulating quantum systems \cite{Lee2023qchemistry}. Similarly, there is an informal consensus that NP-hard optimisation problems will retain their exponential scaling even in a quantum setting \cite{Au-Yeung2023optimisation}.

Careful budgeting is especially crucial when integrating HHL within classical computational frameworks, where the necessary processes of encoding classical data into quantum states and subsequently extracting it post-computation result in critical bottlenecks.
These steps introduce significant computational overheads, which not only diminish the overall efficiency but also render the prospect of obtaining a \emph{useful} exponential advantage extremely difficult for practical configurations. 
 
Of the bottlenecks associated with data transfer between classical and quantum realms, the exhaustive readout required to extract a full solution from quantum computations is particularly troublesome and severely limits the feasibility of achieving net speedups in real-world applications \rev{\cite{aaronson2015read, pareek2024demystifying,jin2022time}}.
Our proposed predictor-corrector offers a novel framework for circumventing this obstacle and leveraging quantum potential in practice. 
\rev{This results in} not just a methodological choice but a conceptual framework for integrating quantum algorithms into broader computational paradigms, thinking of the quantum computer as a specialised processor, which is called as part of a broader algorithm \cite{callison2022hybrid}.
By demonstrating the potential to efficiently extract output from a quantum computation that is meaningful even in a classical context, our predictor-corrector secures a \emph{useful} polynomial advantage over direct application of HHL.
Highlighting the importance of budgeting for data transfers, our work further advocates for a re-imagined approach to applying quantum algorithms to complex challenges, emphasizing a more holistic integration with classical methods while potentially opening up new research avenues and fields of application. 

Next we present some relevant background for the HHL algorithm as well as incompressible smoothed particle hydrodynamics (ISPH) \rev{and other relevant classical methods} that will be used to demonstrate the proposed predictor-corrector. Two forms of the latter, a hybrid and fully quantum predictor-corrector are proposed with their strengths and weaknesses evaluated. 
Despite the primary choice of smoothed particle hydrodynamics here, it is important to note that the methods remain completely general as highlighted with later applications to plasma and reactive flow simulations. 

\section*{Methods}
\subsection*{Solving systems of linear equations}
A Linear Systems Problem (LSP) with $N$ equations and $N$ unknowns is typically formulated as 
\begin{equation}\label{e:1}
    \mathbf{Ax=b},
\end{equation}
with the $N\times N$ matrix of coefficients $\mathbf{A}$, the known vector $\mathbf{b}$ and vector of unknown solutions $\mathbf{x}$. 
If the matrix is invertible, a set of solutions is readily obtained as $\mathbf{x=A^{-1}b}$. Several classical methods for achieving this exist \cite{saad2020iterative} and include direct methods such as Gaussian elimination or LU decomposition as well as Krylov methods such as the conjugate gradient approach \cite{shewchuk1994introduction}. 
In contrast to this, HHL addresses the quantum counterpart to Equation \ref{e:1}, where $A$ is now Hermitian and the vectors $\ket{x}$, $\ket{b}$ are valid (i.e. subject to normalisation) quantum states, 
\begin{equation}\label{e:2}
    A\ket{x}=\ket{b}.
\end{equation}

In the context of this quantum LSP, HHL seeks the quantum state $\ket{x}$. The associated circuit and a more detailed explanation of the process is provided in Supplementary A. Briefly, HHL consists of five main components, namely state preparation, quantum phase estimation (QPE), controlled rotation, inverse quantum phase estimation (IQPE) and measurement. State preparation involves encoding the entries of $\ket{b}$ into the quantum computer, followed by QPE to estimate the eigenvalues of $A$. These eigenvalues are then inverted using controlled rotations where the angle is inversely proportional to the corresponding eigenvalue. This forms an efficient simulation of the effect of $A^{-1}$ on the quantum state without explicit evaluation of the matrix. IQPE is then needed to disentangle the system and measurement of the ancilla determines whether a particular run was successful.

HHL scales as $\mathcal{O}(\text{log}(N)s^2\kappa ^2 /\epsilon)$ for precision $\epsilon$, matrix condition number $\kappa$ and matrix sparsity $s$. Thus, it solves the system in logarithmic time with respect to problem size, exponentially faster than classical methods \cite{harrow2009quantum}. Moreover, later works exponentially improved the dependency on precision \cite{childs2017quantum} and improvements have lead to the development of sparsity independent alternatives \cite{wossnig2018quantum}. The methods proposed in this paper are readily compatible with the different variations and so no further distinctions are made here.
However, achieving this exponential advantage with HHL is contingent on the following caveats \cite{aaronson2015read}: 

\begin{itemize}

    \item The quantum computer must be able to apply unitary transformations of the form $\exp(-iAt)$ efficiently. This usually requires the matrix $A$ to be sparse and well conditioned. The requirement of this matrix being Hermitian is not limiting, as a non-Hermitian matrix can be readily padded into Hermitian form.
    \item The quantum state, $\ket{b}$, needs to be prepared efficiently. This might involve encoding the entries of vector $\mathbf{b}$ into the amplitudes using quantum random access memory (QRAM), or directly constructing the state according to an explicit formula.  
    \item On successful completion of HHL, the final output is $\ket{x}$ and it is assumed that meaningful post-processing can be done efficiently.  
\end{itemize}

It is very challenging to satisfy all of these conditions for practical configurations of interest. Often, when actually including the budgeting for any associated complexities of state preparation and solution post-processing there is no longer any net exponential speedup to leverage. However, we endeavour to tackle some of these bottlenecks in the hopes of \rev{obtaining} a useful polynomial advantage. 
The final caveat, addressed here, is particularly troublesome and commonly referred to as the \emph{readout problem}, in that determining the vector $\ket{x}$ is not directly equivalent to determining its components $x_i$, which are commonly the desired result in classical frameworks. This is a consequence of the inherent quantum nature of Equation \ref{e:2} that encodes a solution in an amplitude basis and is also a crux for many other quantum algorithms. 
HHL is thus fundamentally different to classical methods in that it facilitates effectively finding the state $\ket{x}$, but does not \emph{a priori} guarantee effective readout of the complete solution. There is currently no available approach to extract the complete solution without annihilating any quantum speedup due to the need for repeated measurements. 

\rev{While using HHL to generate replicas of the state $\ket{x}$ has been proven useful as a subroutine for other quantum algorithms \cite{duan2020survey,guan2020hhl}, applicability to classical problems remains challenging. 
Although it is feasible to obtain the expectation value of some operator $M$, $\bra{x}M\ket{x}$ efficiently, the practical utility of this remains unclear as in a classical context it is usually the complete solution that is the desired state.}  

This study introduces a novel approach to tackle this challenge, by adapting the HHL algorithm into a predictor-corrector framework, designed to complement rather than supplant classical computational strategies. 
This approach leverages the algorithm's capacity for sampling from a solution distribution without necessitating complete solution disclosure. In doing so, we refocus the efforts from just computing outputs via HHL to strategically employing these outputs in synergy with classical algorithms. 
This paradigm shift strives towards mitigating the readout problem and enhancing the practical utility of quantum algorithms in solving classical challenges.
To circumvent the limitations of current quantum hardware, the desired quantum state is simulated, as detailed in Supplementary B. This approach is justified since the novelty of our work lies in how to utilize the quantum state once it is obtained, rather than in improving the generation of the state itself. 

\begin{figure*}[h!]
\centering
\begin{subfigure}[b]{0.32\textwidth}
\includegraphics[width=\textwidth,height=5.9cm]{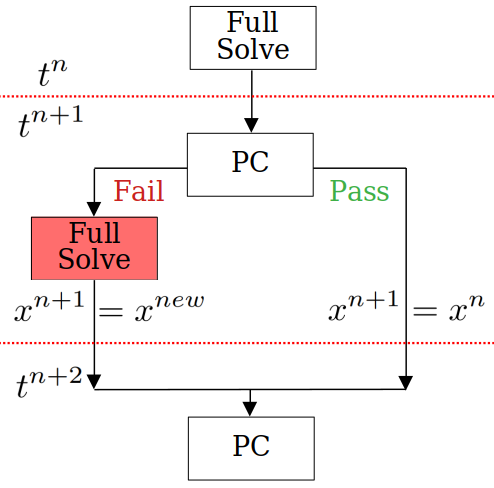}
\caption{}\label{g:1}
\end{subfigure}
\hfill
\begin{subfigure}[b]{0.32\textwidth}
\includegraphics[width=\textwidth]{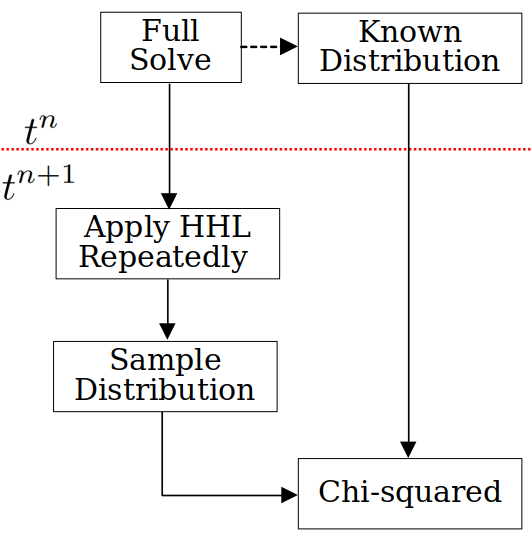}
\caption{}\label{g:2}
\end{subfigure}
\hfill 
\begin{subfigure}[b]{0.32\textwidth}
\includegraphics[width=\textwidth]{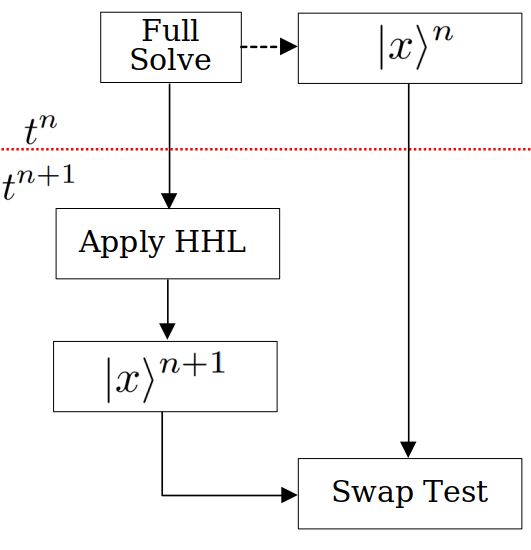}
\caption{}\label{g:3}
\end{subfigure}
\caption{(a) Interface of the proposed predictor-corrector (PC) with the global algorithm. At each time step (separated by red dotted lines), the last known solution is fed into the PC to determine whether a full classical solve is needed at the current time step or if  previous values for the solution can be carried forwards. The PC black box consists of either HHL and a classical statistical test for the hybrid PC (b) or HHL and a swap test for the quantum PC (c). }
\label{fig:methods}
\end{figure*}
\subsection*{A HHL based predictor-corrector}
The role of our predictor-corrector (PC) is to determine how different the solution to a LSP is at the next time step compared to the currently known value. If the difference is greater than some pre-determined rejection threshold, this time step requires an expensive, full classical solve. However, it will be demonstrated that for numerous applications of interest, it is possible to successfully bypass this solve and proceed to the next time step while retaining the previously computed solution as displayed in Figure \ref{g:1}.
To this end, two different approaches are proposed here, namely a hybrid predictor corrector (H-PC) as shown in Figure \ref{g:2} and a fully quantum predictor-corrector (Q-PC) as illustrated in Figure \ref{g:3}. 

In the H-PC scheme, assuming the full state is known at $t^n$, the H-PC consists of applying HHL to the LSP at $t^{n+1}$ for a number of repetitions to obtain a sample distribution. A classical statistical test is then used to judge the likelihood of this sample distribution having the same underlying distribution as one based on the last known solution. Many classical tests exist for this and the simplest option of a Chi-squared test \rev{\cite{stigler2002statistics,franke2012chi}} is used here, although this can be readily replaced with more optimised alternatives. 
The essence of this approach lies in its ability to infer whether the distribution at $t^{n+1}$ has significantly diverged from the known distribution at $t^{n}$. The advantage of integrating the Chi-squared test is that it requires considerably fewer samples compared to the quantity needed to explicitly solve the LSP using HHL. 

Alternatively, the Q-PC stores solutions at $t^n$ and $t^{n+1}$ as quantum states.
These are then used as the inputs for a quantum swap test \rev{\cite{montanaro2013survey,buhrman2001quantum}}, after-which the ancilla bit is sampled to determine the probability of observing the $\ket{0}$ state, as described by  

\begin{equation}\label{e:3}
    P(0)=\frac{1}{2}+\frac{1}{2}|\innerprod{x^n}{x^{n+1}}|.
\end{equation}

Thus, the probability is directly related to the degree of overlap as embodied by the inner product. 
This allows pre-determining a minimum probability, above which the degree of overlap is deemed sufficient to warrant skipping the full classical update. 
The elegance of this approach is that the sampling has been reduced to that of a single (ancilla) bit, independently of problem size, as opposed to sampling the full solution state space in the H-PC. 
This comes with the additional computational cost of executing a swap test and also requires storage of two independent quantum states in memory. \rev{However, this is not a significant drawback even in the context of near term hardware.}

Each predictor-corrector scheme offers user-defined parameters, allowing for preemptive adjustments to modulate the frequency of \rev{LSP} bypasses (also referred to as \emph{skips}).
The H-PC approach includes determining the number of samples generated via HHL for the Chi-squared test, alongside a critical P-value. The sample count influences the reliability of the judgement, while the P-value establishes the confidence level for rejecting (or upholding) the null hypothesis. This P-value signifies the probability of encountering a test statistic as extreme as, or more so than, what the test data yielded, presuming the null hypothesis is valid. Typically, a high P-value aligns with the null hypothesis, indicating a high likelihood of observing such statistics, whereas a low value indicates a low likelihood, suggesting rejection of the null hypothesis. Thus, users can set a predefined critical threshold, beneath which the outcomes are considered significantly divergent to warrant the null hypothesis's rejection.
Given the null hypothesis postulates no statistical discrepancy between the sampled distribution and the expected distribution from the last known time step, rejecting it necessitates a full pressure update at the current time step.
The mechanism for the Q-PC is straightforward, governed by a critical probability threshold. Beyond this threshold, the extent of overlap between states is deemed sufficient to forego the pressure update, with this critical probability determined by a Equation \ref{e:3} once the user sets an acceptable inner product threshold.

\subsection*{Incompressible Smoothed Particle Hydrodynamics}
ISPH is a popular particle based, Lagrangian flow solver with applications across a wide range of disciplines \cite{ellero2007incompressible,morris1996analysis}. 
For a more complete discussion, the interested reader is referred to the many excellent reviews on the topic \cite{lind2020review,monaghan2012smoothed}. 
In the scope of this work, it will suffice to appreciate that at each time step a collection of equations are solved that compute particle accelerations, update velocities and subsequently advance particle positions in space as outlined in Supplementary C. 
\rev{Of particular importance is that in order to enforce incompressibility the following Poisson equation for pressure is solved,}  
\begin{equation}\label{e:99}
    \nabla \cdot \left( \frac{1}{\rho} \nabla P \right)_i = \frac{1}{\Delta t} \nabla \cdot \mathbf{u}_i^* ,
\end{equation}
\rev{with $P$, $\rho$ and $\bold{u}_i^*$ referring to pressure, density and intermediate particle velocity respectively. When discretised by the ISPH operator, Equation \ref{e:99} becomes a linear system of equations.}  
This LSP is traditionally solved at every time step to compute the pressure correction needed to provide a divergence free velocity field. 
The authors propose that for many flows of interest this pressure correction does not need to be evaluated at every time step and proceed to evaluate this claim using the novel PC. 

Furthermore, SPH employs kernels with compact support (i.e. particles cannot influence particles that are far away). As a result, the matrix for this LSP is sparse and large. In addition, it forms a key bottleneck for classsical ISPH, making it an ideal candidate for the proposed PC strategy. 

\rev{\subsection*{Other Methods}}
The Vlasov-Poisson system of equations is fundamental for describing the evolution of particles in a self-consistent field, with applications spanning plasma physics \cite{han2015stability}, astrophysics \cite{rein2007collisionless}, and nuclear dynamics \cite{reinhard1995dissipative}. In plasma physics, the Vlasov equation \cite{feix2005universal} captures the evolution of the plasma distribution function in multidimensional phase space. In the absence of strong magnetic fields, this is coupled with the Poisson equation which defines the relationship between the electrostatic potential ($\Phi$) and charge density ($q$), as shown in Equation \ref{e:5}. 
\begin{equation} \label{e:5}
    \nabla^2 \cdot \Phi = \frac{q}{\epsilon_o}.
\end{equation}
Discretisation of this Poisson equation results in another sparse LSP. 

Alternatively, in reactive flow simulations the Navier-Stokes are supplemented with an equation for the reactive scalar of the form 
\begin{equation}\label{e:6}
\frac{\partial Y_k}{\partial t} + \nabla \cdot (\mathbf{u} Y_k) = \nabla \cdot (D_k \nabla Y_k) + R_k ,
\end{equation}
where $Y_k$ represents the mass fraction of the $k$-th species, with velocity $\mathbf{u}$, some diffusion coefficient $D_k$ and the reactive source term $R_k$. 
The reactive source term for a particular species is a function of other species in the system and is generally solved as a separate system of ordinary differential equations (ODEs), prior to inclusion in the global transport equation. 
It is possible to linearise this set of ODEs into a LSP, as demonstrated by recent attempts to solve similar systems using HHL \cite{akiba2023carleman,becerra2022quantum}. 

\section*{Results and Discussion}

As an initial proof of concept, we first apply our predictor-corrector to the Taylor-Green Vortex (TGV), a canonical benchmark configuration for validating computational methods in the realm of classical fluid dynamics. Following this, we simulate the 2D dam break scenario, involving a vertical column of fluid initially at rest that is released to form waves, demonstrating the intelligent adaptability to complex flow dynamics of the proposed predictor-corrector. Finally, we explore other potential applications, in the realms of plasma physics and reactive flow simulation specifically, by addressing the two-stream instability and flame-vortex interaction cases respectively.

\subsection*{Taylor-Green Vortex}
\begin{figure*}[h!]
\centering
\hfill
\begin{subfigure}[b]{0.44\textwidth}
\includegraphics[width=7cm,height=7.2cm]{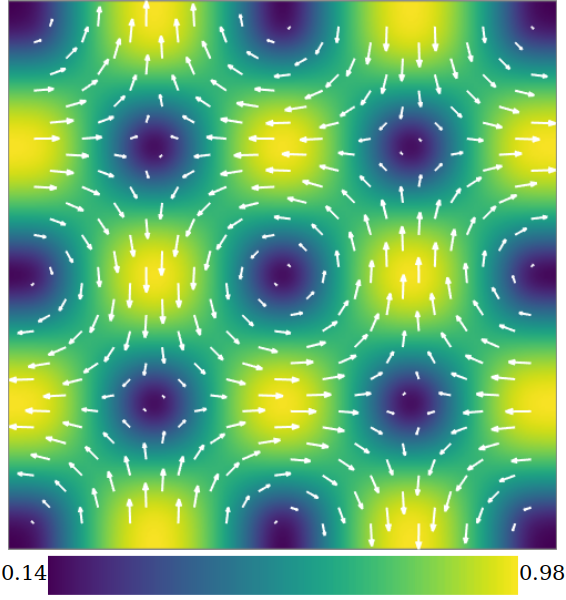}
\caption{}\label{p:1}
\end{subfigure}
\begin{subfigure}[b]{0.54\textwidth}
\includegraphics[width=\textwidth]{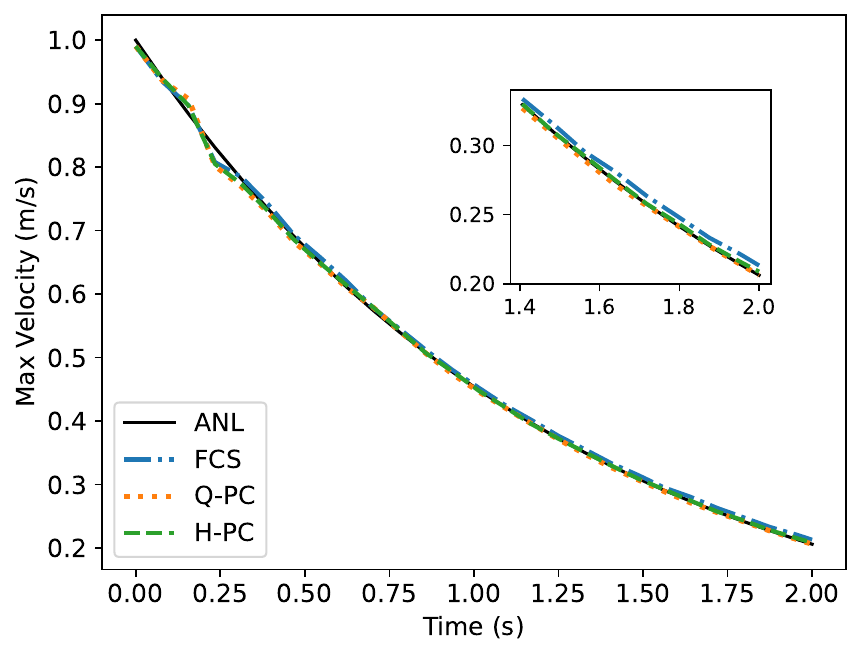}
\caption{}\label{p:2}
\end{subfigure}
\caption{(a) Snapshot of the domain for the Taylor-Green vortex (TGV), with vector arrows representing velocity to highlight the lattice of decaying vortices. (b) Decay profile of max velocity for the analytical (ANL), full classical solve (FCS), quantum predictor-corrector (Q-PC) and hybrid predictor-corrector (H-PC) solutions. }
\label{fig:1}
\end{figure*}

The TGV exemplifies a quintessential flow pattern consisting of a periodic array of counter-rotating vortices, with a typical flow profile depicted in Figure \ref{p:1}. Despite its apparent simplicity, this fundamental flow can offer an idealised environment to study complex fluid dynamics such as vortex stretching and the eventual transition to turbulence. 
The following ISPH simulations consist of $32^2$ particles in two dimensions with a Reynolds number of $100$ in order to investigate any impact of skipping resolution of the Poisson equation, while still in the well understood laminar regime. 

Figure \ref{p:2} showcases the exponential decay of velocity in the domain for the analytical (ANL), full classical solve (FCS) i.e. Poisson equation is solved directly at every step, as well as the quantum (Q-PC) and hybrid (H-PC) predictor-corrector simulations.
Both the hybrid and quantum PCs skip updating pressure for approximately $45-50\%$ of the time steps and still retain excellent agreement with the global solution. The resultant decay profiles are almost indistinguishable with analytical predictions. 
This validates the fundamental motivation behind the predictor-corrector framework, by demonstrating that it is indeed possible to skip the computationally expensive update of pressure. Moreover, this can be done for a substantial proportion of time steps, without excessive deterioration of the global solution. 
Although there is a noticeable dip in the early stages, this is true for the FCS as well and is thus not an artifact of the proposed method but rather of SPH itself. Moreover, at later times the PC simulations seem to agree with the predicted values better than when using a FCS at the same resolution, as illustrated in the focused inset. This suggests that for some configurations, repeated updates of the pressure correction term may be unnecessary and having an intelligent tool to detect this is very likely to be useful. 

\begin{figure*}[h!]
\centering
\includegraphics[width=0.5\textwidth]{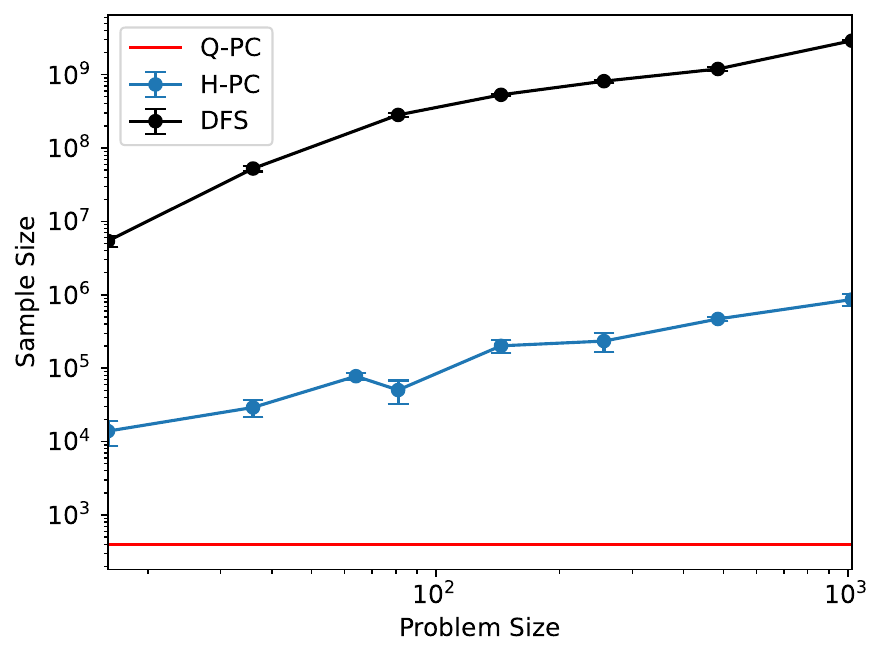}
\caption{Scaling of the required number of samples with problem size in order to measure the desired output. This includes when using HHL to obtain a direct full solution (DFS) i.e. each of the probability amplitudes, as well as requirements for the quantum/hybrid predictor-correctors (H-PC/Q-PC). }
\label{p:sclaing}
\end{figure*}

\begin{figure*}[h!]
\centering
\hfill
\begin{subfigure}[b]{0.49\textwidth}
\includegraphics[width=9cm,height=6.1cm]{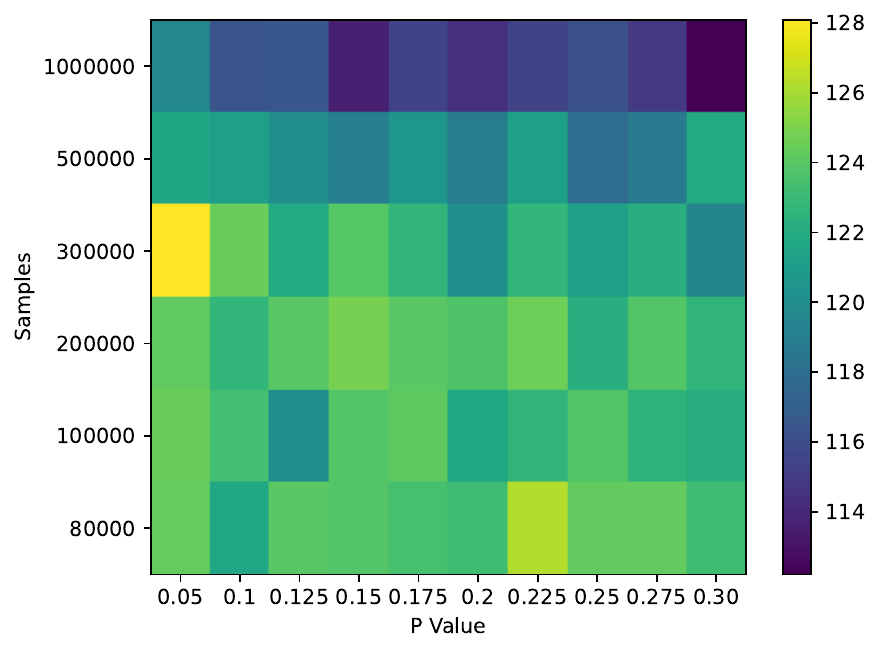}
\caption{}\label{p:3}
\end{subfigure}
\begin{subfigure}[b]{0.49\textwidth}
\includegraphics[width=\textwidth]{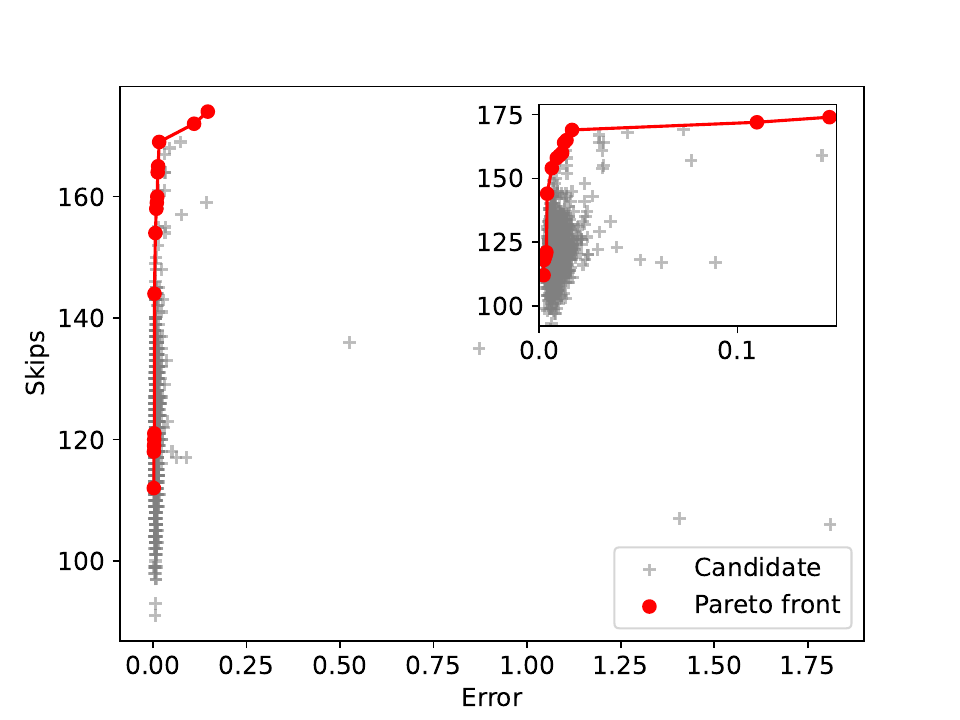}
\caption{}\label{p:4}
\end{subfigure}
\caption{(a) Heat map colorized to represent the number of skips for particular combinations of sample size and P-Value. Results were averaged across 25 repetitions for each configuration. (b) Approximate Pareto front when treating each simulation as a multi-objective optimisation problem, seeking to minimize error and maximise skips. Pareto inferior combinations are included in grey as candidates and an inset that ignores outliers is provided.}
\end{figure*}
Having demonstrated the potential for accelerating the global algorithm by skipping resolution of the Poisson equation, we next explore the trade-offs involved. This exploration is visualized in Figure \ref{p:sclaing}, which contrasts the requisite number of samples for obtaining a comprehensive solution via HHL against that required by the proposed PC strategies. 
As anticipated, aiming for a direct full solution (DFS) through HHL leads to sample size scaling steeply with problem size, as approximately $N^{1.4}$, although systematic deviations from linearity \rev{(in log-log scale)} are visible, suggesting \rev{a more complicated dependency on problem size such as additional logarithmic factors for example.} 
Once included in the overall cost budget, this would severely erode any exponential advantage intrinsically offered by HHL. Moreover, the high baseline cost at each problem size severely limits the range of problem sizes that can be investigated, especially in the context of near term quantum hardware. 

Conversely, the Hybrid Predictor-Corrector (H-PC) markedly reduces the scaling dependency, achieving a near-linear scaling ($N^{1.0}$), while simultaneously reducing the baseline cost by multiple orders of magnitude. 
This is further consolidated by the Q-PC which achieves near constant scaling due to the measurements being applied to only a single ancilla qubit, independently of problem size. 
These finding underscore the core contribution of this work, demonstrating that by designing innovative applications for HHL, it is possible to provide meaningful output even in a classical framework, while still optimising global efficiency and practical utility. 

For determining the complete solution, the necessary number of samples was evaluated based on achieving an approximation of the true solution norm, as found by classical methods, within a $5\%$ error margin for a single time step. However, the H-PC necessitated a different criterion due to the distinct lack of a final discrete solution vector. This criterion was established as the requisite number of samples to identify a statistical divergence between two states, with a significance level (P-value) of less than 0.05. The comparison involved solution vectors separated by ten time steps, a decision heuristically motivated by the insight that such substantial intervals between pressure updates lead to simulation instability for the TGV. Thus, a proficient PC strategy should reliably discern statistical disparities and mandate a comprehensive pressure update for such a gap. \rev{For the Q-PC, taking each ancilla bit measurement as an independent Bernoulli trial and then approximating the Binomial distribution with a normal distribution, it is possible to rearrange the expression for margin of error into,}

\begin{equation}
    n=\frac{Z^2 p(1-p)}{E^2},
\end{equation}
\rev{where $n$ is the required number of samples, $Z$ refers to the Z-score for a particular confidence interval, $p$ is the probability of measuring $0$ for the ancilla and $E$ is the acceptable error margin. The worst case scenario of $p=0.5$ was then used in conjunction with a $95\%$ confidence interval and $5\%$ error margin.} 

Figure \ref{p:3} demonstrates how variations in user-defined parameters affect skip frequencies for the TGV, averaging results over 25 iterations for each setup. 
Notably, by definition setting a higher P-value threshold inherently increases the risk of erroneously dismissing the null hypothesis, this manifests as a lower frequency of skips. Consequently, the highest frequency of skips is typically observed for the lowest P-value. This can be accompanied with some degree of instability, where there is a small but non-zero chance of the simulation terminating early due to large oscillations, unless using a very large sample size.  
Similarly, as indicated by the collection of lower skips at the top of Figure \ref{p:3}, increasing sample size generally decreases skips, likely due to a more representative sample improving the detection of statistical anomalies.

However, these trends are not strictly linear and may include some localised deviations. Adjusting parameters to increase skip frequency does not always produce a directly proportional rise in skips, likely due to the intricate interplay between the PC scheme and underlying fluid dynamics. For example, introducing a higher number of initial skips initially can result in increased flow fluctuations, which may subsequently require increased pressure recalculations. Nevertheless, precisely defining this correlation is not entirely relevant as the key insight here is that variations in skip count tend to be minor, suggesting that the method is resilient to changes in parameter settings.

This is further demonstrated by exploring the problem state space and approximating the Pareto front as done in Figure \ref{p:4}.
This approach frames the task as a multi-objective optimization problem, aiming to maximize the number of skips while concurrently minimizing error. 
The tight clustering of points in the low error region close to the Pareto front suggest that large oscillations are rare. 
Indeed significantly large errors occur for only approximately 4 candidate solutions out of 1500 total simulations (i.e. 25 iterations each for 60 combinations of sample size and P-value). Additionally, the focused inset highlights a significant advantage of the proposed PC algorithm, where a notably sharp increase in skips can be achieved initially with only a marginal elevation in error, as depicted by the steep initial slope of the Pareto curve.
This trend, eventually approaches a limit beyond which further skips result in relatively larger errors, evidenced by the curve's eventual plateau. However, an intelligent predictor-corrector allows users to navigate this Pareto front, leveraging the inherent trade-offs of the problem to optimize performance effectively.
The Q-PC scheme was found to be similarly robust, and will be the preferred method of choice henceforth for subsequent examples due to the improved scaling characteristics. 

\subsection*{Dam Break}
\begin{figure*}[h!]
\centering
\hfill
\begin{subfigure}[b]{0.24\textwidth}
\includegraphics[width=\textwidth]{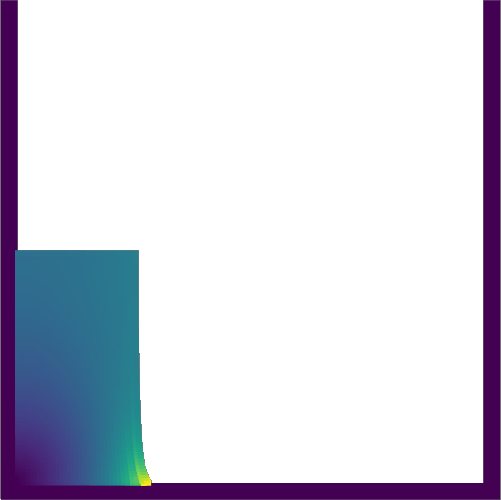}
\caption{$T_1$}
\end{subfigure}
\begin{subfigure}[b]{0.24\textwidth}
\includegraphics[width=\textwidth]{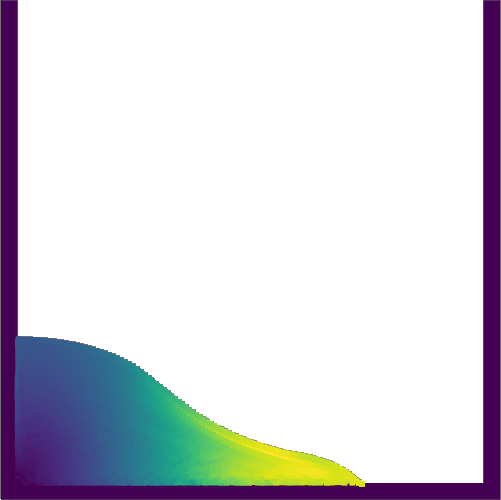}
\caption{$T_2$}
\end{subfigure}
\begin{subfigure}[b]{0.24\textwidth}
\includegraphics[width=\textwidth]{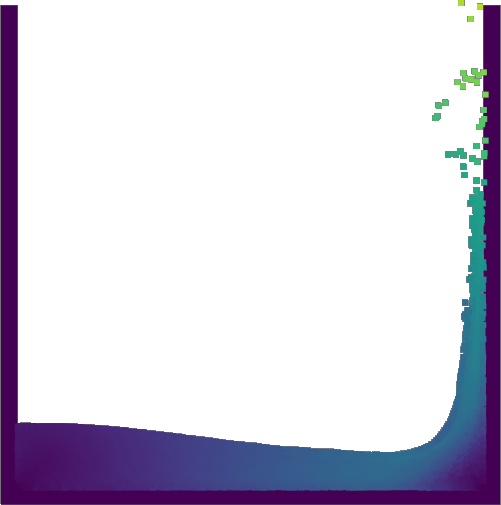}
\caption{$T_3$}
\end{subfigure}
\begin{subfigure}[b]{0.24\textwidth}
\includegraphics[width=\textwidth]{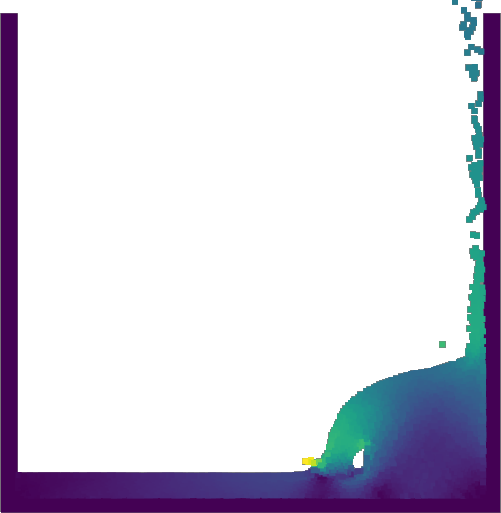}
\caption{$T_4$}
\end{subfigure}
\caption*{FCS}
\centering
\hfill
\begin{subfigure}[b]{0.24\textwidth}
\includegraphics[width=\textwidth]{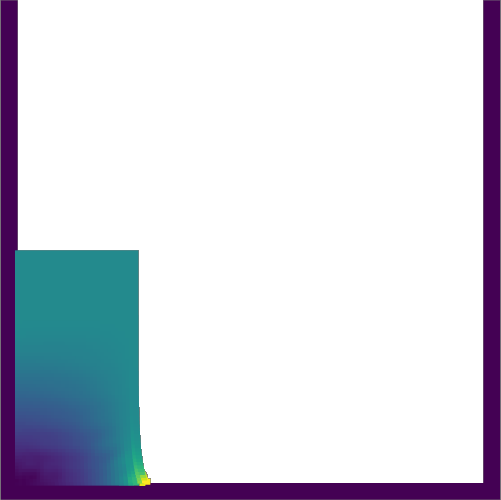}
\caption{$T_1$}
\end{subfigure}
\begin{subfigure}[b]{0.24\textwidth}
\includegraphics[width=\textwidth]{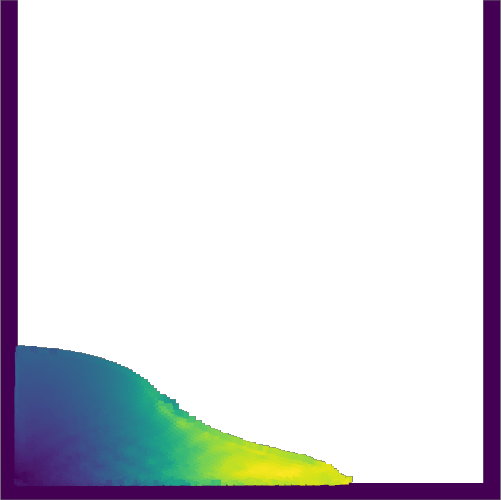}
\caption{$T_2$}
\end{subfigure}
\begin{subfigure}[b]{0.24\textwidth}
\includegraphics[width=\textwidth]{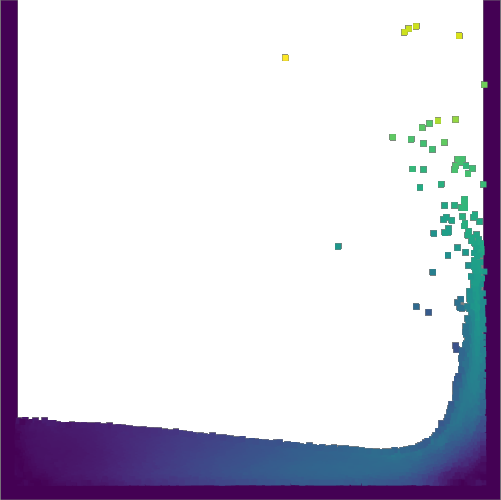}
\caption{$T_3$}
\end{subfigure}
\begin{subfigure}[b]{0.24\textwidth}
\includegraphics[width=\textwidth]{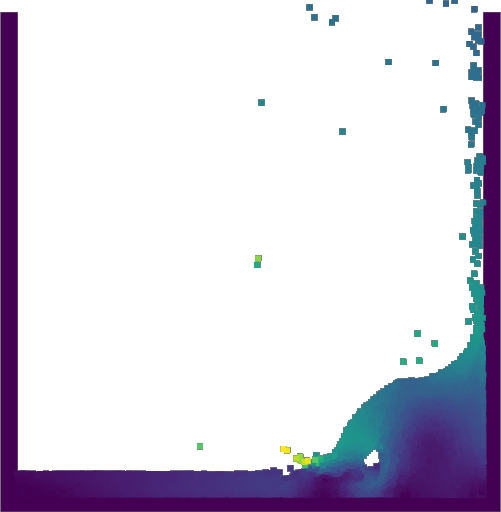}
\caption{$T_4$}
\end{subfigure}
\caption*{Q-PC}
\begin{subfigure}[b]{0.49\textwidth}
\includegraphics[width=\textwidth]{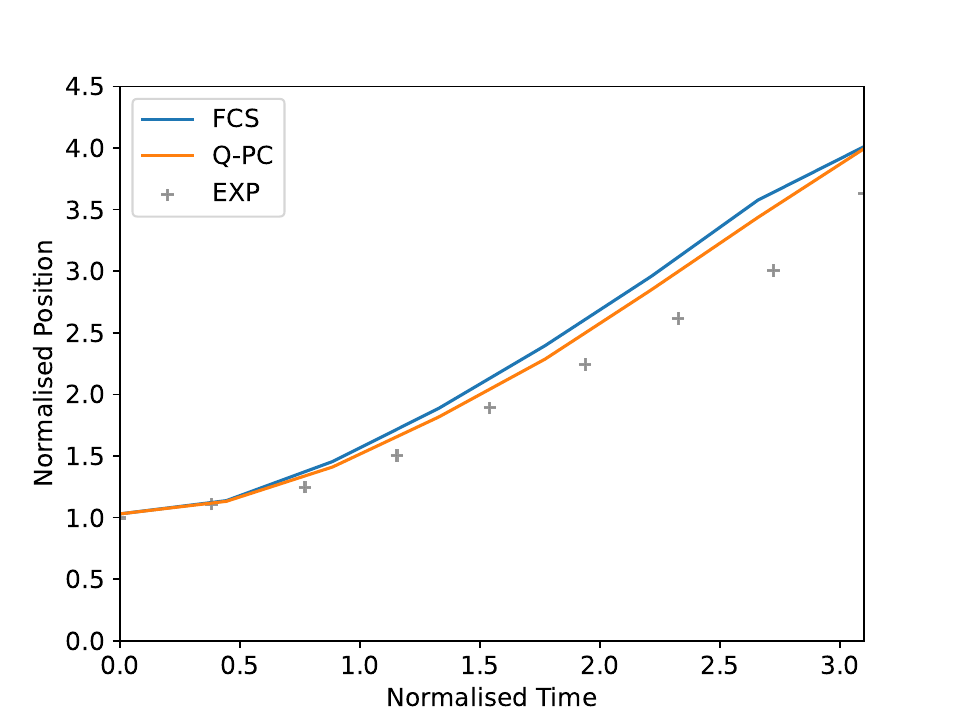}
\caption{}\label{p:6}
\end{subfigure}
\begin{subfigure}[b]{0.49\textwidth}
\includegraphics[width=\textwidth]{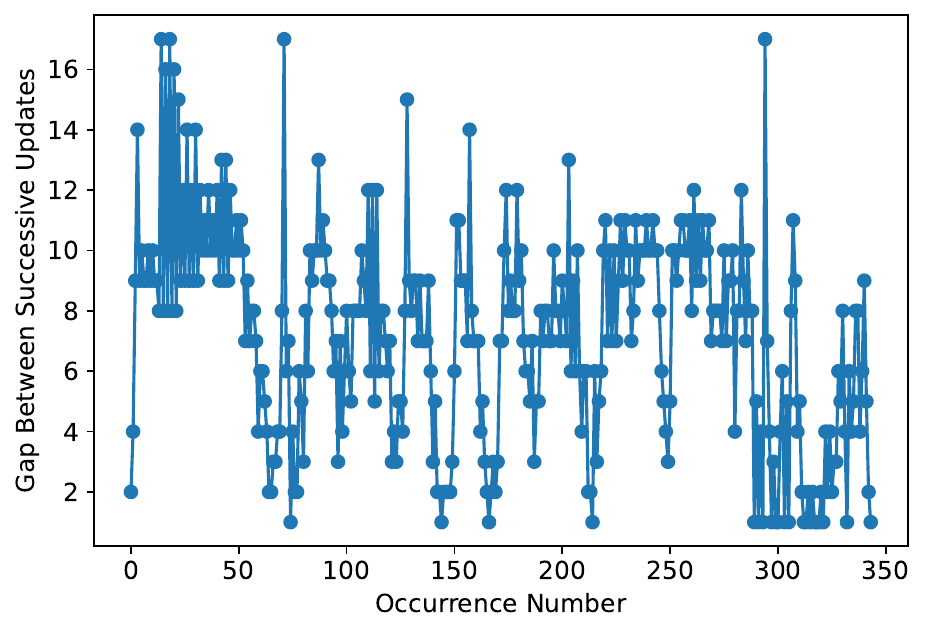}
\caption{}\label{p:7}
\end{subfigure}
\caption{(a-d) Collapse of a water column in a tank simulated using a full classical solve (FCS) that resolves a pressure-Poisson equation at every time step. (e-h) Collapse of a water column in a tank using a quantum predictor-corrector (Q-PC) to resolve a pressure-Poisson equation for only $16\%$ of the time steps. (i) Temporal evolution of the leading wave edge in space following collapse of the water column. Results for a full classical solve (FCS), quantum predictor-corrector (Q-PC) and experiments \cite{koshizuka1996moving}(EXP) are included. (j) Variation in the number of successive skips of the pressure-Poisson equation between consecutive full pressure updates, as required by the Q-PC.}
\label{p:5}
\end{figure*}

Furthermore, the proposed PC is about more than just the quantity of skips, but also takes into account their intelligent allotment throughout the simulation. This is demonstrated here using a common real world application for SPH, known as the dam break problem, as illustrated in Figure \ref{p:5}. The configuration consists of a a vertically stacked column of fluid initially held at rest in tank. At the start of the simulation, the barrier restraining the fluid column is removed leading to its subsequent collapse, prior to wave formation and propagation through the domain. This is a commonly used benchmark in the SPH community due to how effectively it assesses a solver's capability to accommodate free surface flows, boundary interactions and complex fluid interactions. Moreover, the scenario effectively tests the numerical stability of the method under rapid changes in fluid velocity and pressure, as well as its convergence towards a physically accurate solution.

The simulations depicted in Figure \ref{p:5} utilize a total of 
$2278$
 fluid particles and $1660$ boundary particles, summing up to $3938$ particles. These snapshots illustrate the temporal evolution of the fluid dynamics as modeled by both the FCS and Q-PC algorithms. It is readily apparent that the Q-PC algorithm proficiently captures the essential phenomenological behaviors of the flow. This includes the adept resolution of complex dynamics such as wave formation and propagation, along with the wave's impact against the wall and its subsequent rebound.
The slight variations in the positions of some individual particles between the simulations are of minimal concern, especially considering that particle shifting is a common practice in classical SPH to enhance numerical stability. 
The primary focus lies in the overall accuracy of the flow dynamics captured by the simulations, which is done by averaging and not by the exact positions of individual particles. 

The ability of the Q-PC algorithm to accurately model these critical aspects of the dam break problem are further demonstrated in Figure \ref{p:6}, where despite only updating the pressure $16\%$ of the time, the wave front is excellently resolved throughout the simulation with minimal differences with the FCS. Note that although the experimental results have been included here for completeness, it is well known that SPH simulations tend to over-predict the edge position due to friction between the leading edge and bottom wall leading to lower velocities in experimental settings. Thus, the minimal deviation between Q-PC and FCS is well within acceptable bounds.

When the same number of skips are evenly distributed in the same configuration, this leads to simulation failure, triggered by the introduction of instabilities in the flow.
This underlines a critical advantage of the proposed PC approach, not only does it permit bypassing the computationally intensive pressure solve, but it does so in a manner that takes into account the intrinsic dynamics of the problem at hand.

This principle is further illustrated in Figure \ref{p:7}, which depicts the distribution of skips between successive updates throughout the simulation, encompassing the approximately 350 updates in total. Initially, as the bulk of the fluid column remains static with only the leading edge beginning to detach, this phase is marked by a significant interval between updates, denoted by successive gaps in the high range of 10-14. This reflects the relative stability of the majority of the fluid during the early stages. As the wave starts to collapse, the frequency of necessary updates increases, particularly observed in the later stages where the number of skips between updates consistently falls within the 1-2 range. This adjustment corresponds to the wave's impact and rebound against the wall, a phase where the entire domain experiences substantial and swift changes. Moreover, the presence of several points along the length of local peaks and troughs in the skip distribution illustrates a careful, incremental adjustment to the gap between updates, tailored to the fluid's dynamic response. This approach avoids abrupt changes that could compromise simulation stability, reflecting a nuanced adaptation to flow dynamics. 

The adaptive nature of skips within the PC framework, therefore, is not arbitrary but is intricately linked to the fluid dynamics being simulated. By aligning skips with the fluid's behavior, the PC method not only enhances computational efficiency but also maintains the simulation's stability and fidelity, especially during critical phases of large, dynamic changes. 

\subsection*{Other Applications}

\begin{figure*}[h!]
    \begin{subfigure}{0.33\textwidth}
        \centering
        \includegraphics[width=\linewidth]{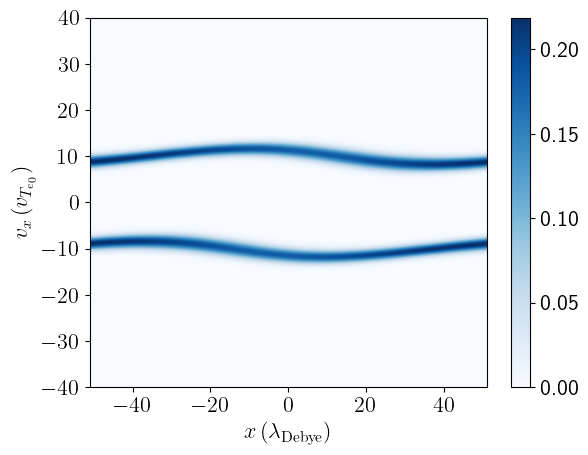} 
        \caption{}
        \label{s:1}
    \end{subfigure}%
    \begin{subfigure}{0.33\textwidth}
        \centering
        \includegraphics[width=\linewidth]{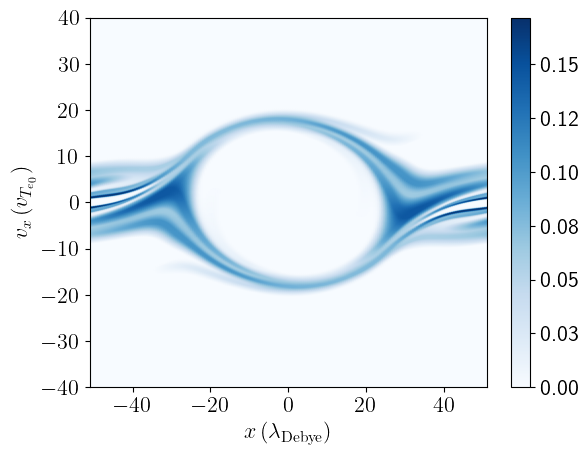}
        \caption{}
        \label{s:2}
    \end{subfigure}%
    \begin{subfigure}{0.33\textwidth}
        \centering
        \includegraphics[width=\linewidth]{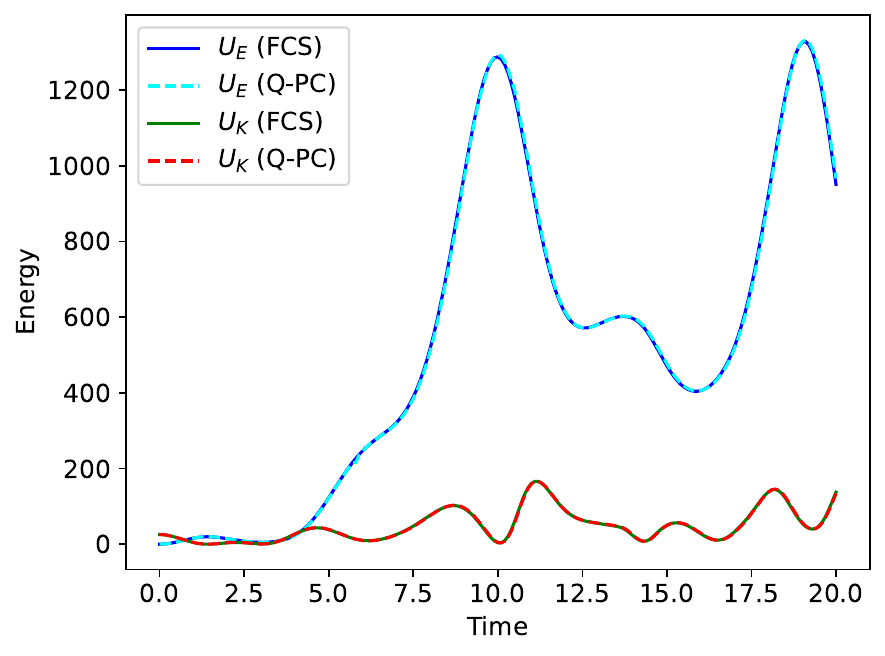}
        \caption{}
        \label{s:3}
    \end{subfigure}
    \centering
  \begin{subfigure}{0.5\textwidth}
        \centering
        \includegraphics[width=\linewidth]{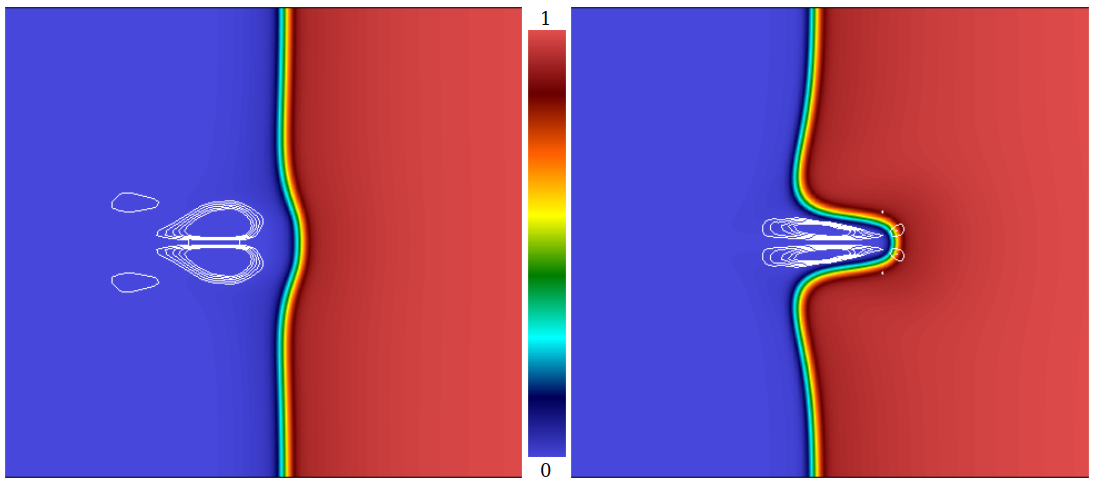}
        \caption{}
        \label{s:4}
    \end{subfigure}%
    \begin{subfigure}{0.25\textwidth}
        \centering
        \includegraphics[width=\linewidth,height=3.9cm]{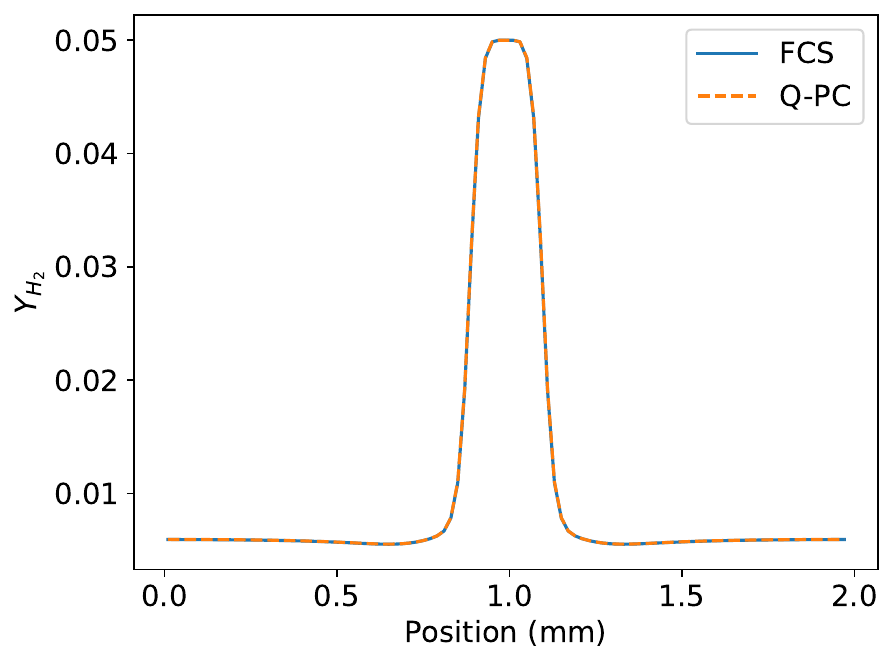}
        \caption{}
        \label{s:5}
    \end{subfigure}%
    \begin{subfigure}{0.25\textwidth}
        \centering
        \includegraphics[width=\linewidth,height=3.9cm]{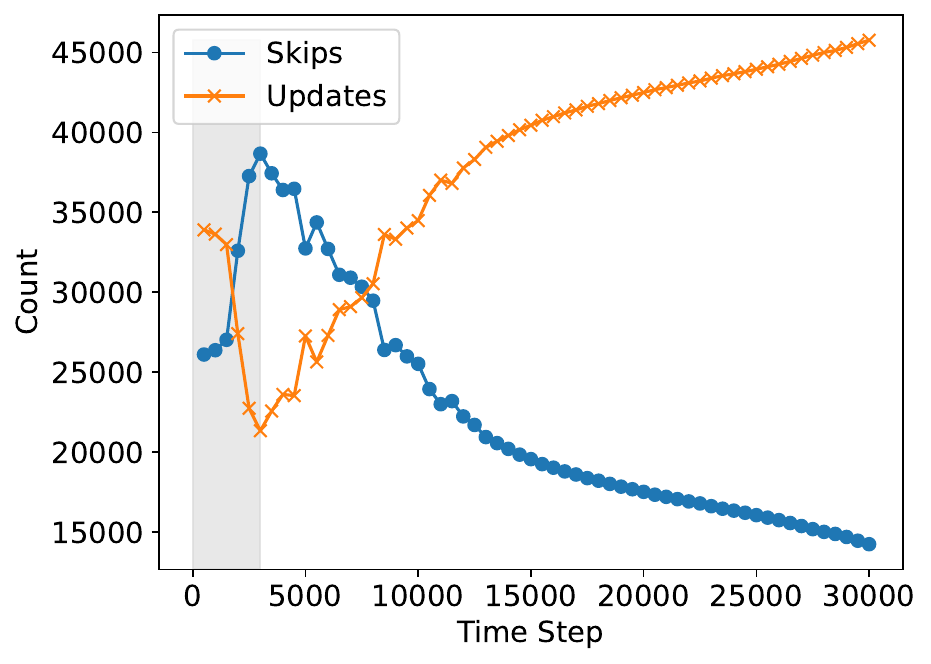}
        \caption{}
        \label{s:6}
    \end{subfigure}   
    \caption{The proposed predictor-corrector can also be applied to plasma (\textbf{a-c}) and reactive flow (\textbf{d-f}) simulations. (\textbf{a-b}) temporal evolution of the two stream instability configuration  as viewed using the plasma electron distribution function in phase space. Initially, the distinct separation in velocity space (i.e. y-axis) described two counter-flowing streams of charged particles. Over time the velocity space merges into a complex wave like pattern due to instability growth. (\textbf{c}) Profiles of plasma electron kinetic energy ($U_K$) as well as total electrostatic energy ($U_E$) for the Q-PC and FCS.
    (\textbf{d}) Temporal evolution of a pair of counter-rotating vortices, as outlined in white iso-contours of vorticity, impacting on a flame front depicted using progress variable (0 in fresh region and 1 in burnt zone). (\textbf{e}) Profiles of hydrogen mass fraction projected along the y-axis for the Q-PC and FCS. (\textbf{f}) Temporal evolution of the ratio of skips to updates in the spatial domain, with an initial window where the flame was still responding to approximate initial conditions greyed out.}
    \label{fig:comprehensive}
\end{figure*}
Moreover, the proposed framework remains completely general, with a versatility that facilitates direct integration into a large host of other classical paradigms. This extends beyond just SPH, including other incompressible approximations like the incompressible Navier-Stokes simulations as well, which similarly aim to maintain divergence-free flow fields. Furthermore, the approach is not limited to pressure solutions alone.
For example, in the domain of plasma physics, the PC methodology could exploit the disparity in timescales between the rapid kinetic dynamics governed by the Vlasov equation and the comparatively slower evolution of the electrostatic field as detailed by the Poisson equation.

Potential application configurations include quasi-neutral plasma, characterized by small net charge densities spread across extensive spatial domains, leading to only a gradually varying electric field. Alternatively, situations with small Debye lengths could benefit from this approach. The Debye length is representative of the distance over which electric fields are screened in plasma. In systems where the Debye length is significantly smaller than the overall system size, electric field variations on scales surpassing the Debye length tend to occur at a slow pace. In such configurations, not enforcing field updates at every time step may suffice to capture the essential underlying dynamics.
As an  initial proof of concept, Figures \ref{s:1}-\ref{s:2} illustrate the temporal evolution of the canonical two stream instability configuration, simulated using  the ESVM \cite{touati2021esvm} example suite.
The interaction between two counter-flowing streams of particles and the background field is seen to manifest as the expected amplification of wave like instabilities. Figure \ref{s:3} consolidates how despite skipping update of the Poisson equation approximately $60\%$ of the time, this had a negligible effect on the global solution. This is true even when the Poisson equation is solving for different physical mechanisms (electrostatics) as opposed to the other coupled equations (plasma distribution functions).  
 
\rev{
As a second example, consider the linearisation of the reactive source term of Equation \ref{e:6}, as detailed in \cite{akiba2023carleman,becerra2022quantum}.
This formulation leads to a distinct LSP at each grid point, representing the set of ODEs for the source term of each species. This contrasts with approaches like ISPH, where the LSP addresses a singular variable (e.g., pressure) across the entire domain. 
This distinction could allow the PC strategy to exploit differences in spatial scales, as opposed to the temporal scales, which have been the focus so far. In typical simulations of premixed combustion, the impact of $R_k$ is predominantly confined to the narrow reaction zone, a small fraction of the overall domain. Consequently, the PC methodology can assess whether significant changes at specific grid points warrant the computation of $R_k$ between time steps.}

A archetypal configuration featuring two counter-rotating vortices interacting with a premixed hydrogen flame front, as simulated using CompReal \cite{rathore2023flame}, serves to illustrate this in Figure \ref{s:4}. Only a tiny fraction of the computational domain, demarcated by values of progress variable between $0.1-0.9$, is chemically active and contains strong gradients in the species. An intelligent predictor-corrector would ensure that the source term is only updated at every time step in this region and not elsewhere where the net impact would be insignificant due to negligibly slow chemical activity.  

Without considering the specifics of linearization techniques here, the demonstration highlights the potential of evaluating the inner product between successive solutions of $R$ to judiciously bypass the update of the overall transport equation at the majority of grid points. This is seen to have negligible effects on spatial mass fraction distributions as shown in Figure \ref{s:5}. Moreover, as the vortex interacts with the flame front and induces local wrinkling, this causes an increase in the total flame surface area. The intelligent PC is able to accommodate this by dynamically increasing the proportion of updates to resolve the net increase in reaction zone in time as illustrated in Figure \ref{s:6}. 
Given the inherent stiffness of ODEs, particularly when employing traditional Arrhenius expressions for reaction rates, calculating $R_k$ becomes notably resource-intensive. Furthermore, the complexity escalates with detailed chemical kinetics, where the number of species—and, by extension, the LSP size—can reach several hundreds or thousands. Hence, the ability to selectively omit this calculation for a substantial portion of the domain emerges as a critical advantage, underscoring the potential of our PC approach in enhancing computational efficiency in chemically reactive simulations.

\vspace{-\baselineskip}
\section*{Conclusions}
Quantum algorithms such as HHL, herald a promising avenue for achieving exponential speedup in solving LSPs. This potential, however, hinges on effectively overcoming the readout problem, a significant hurdle that necessitates multiple samples to extract a comprehensive solution. This work introduces a novel adaptation of HHL into a predictor-corrector framework, effectively bypassing the readout challenge while delivering outputs that are compatible and meaningful within a classical context.

The H-PC approach not only surpasses the conventional application of HHL in terms of scalability but also establishes a foundation for the Q-PC strategy, which boasts scalability that is independent of problem size. This advancement does introduce the trade-off of requiring additional quantum memory capacity to simultaneously manage two states, as opposed to the single-state requirement of the H-PC. Nonetheless, both paradigms represent intelligent methods to leverage the inherent separation of scales characteristic of many classical problems, thereby enhancing overall computational efficiency. \rev{Moreover, the proposed PCs do so with an in-built adaptability that facilitates a dynamic response to changes in the underlying flow dynamics. This is true even for complex flows where attempting to randomly allocate the same number of LSP bypasses instead leads to simulation failure. }

Furthermore, the strategies outlined herein serve as an initial proof of concept, with the potential for refinement and expansion. For instance, within the context of ISPH, the methodology could evolve to base predictions on the rate of change and corrections on prior adjustments, diverging from the current practice of basing predictions on total change and corrections on new adjustments. Alternatively, a more nuanced metric than the Chi-squared test could be readily integrated into the H-PC. \rev{ISPH codes can potentially be dedicating $47\%$ of the total compute time purely to the pressure-Poisson subroutines \cite{guo2018new}, therefore being able to bypass this bottleneck is extremely valuable. }

\rev{To the best of our knowledge, there is currently no existing classical paradigm that offers a similarly intelligent and effective predictor-corrector strategy. However, having demonstrated the viability of bypassing key steps in classical algorithms while preserving solution quality, it is hoped this work will result in future quantum-inspired algorithms that could also perform these tasks efficiently. Specifically, machine learning based classical alternatives could be a potentially rich direction for future work.}
 
The potential for improvements through such modifications invites further investigation, underscoring the dynamic and evolving nature of the framework's potential application. It is hoped that by demonstrating the generality of the approach across fields including fluid dynamics, plasma physics and chemical kinetics, this work will help inspire a wider engagement in the community with regards to how we actually apply quantum algorithms, in addition to designing them. 
\rev{In particular, it is envisaged that this work will highlight how important it is to carefully budget for data transfer costs when applying quantum algorithms to classical problems.  
In order to truly leverage a quantum advantage in practice there remains much work to be done regarding state preparation. Although for some cases this can be done efficiently \cite{holmes2020efficient,zhang2022improved}, in general this remains a key bottleneck.} 

Moreover, in the broader context of quantum computing in the foreseeable future, there is a prevailing view that quantum computers are not destined to supplant classical computers entirely. Instead, the future likely holds a symbiotic coexistence, favouring the emergence of heterogeneous high-performance computing ecosystems that integrate both quantum and classical resources. Algorithms like the proposed PCs are exceptionally well-suited to thrive in such environments due to their intrinsically hybrid design, \rev{as outlined in Supplementary D}. For instance, the PC aspect could be offloaded to quantum computers, operating in parallel and asynchronously from the classical simulation that manages the remainder of the SPH equations. This would allow the classical simulation framework to advance without the need for pressure calculations by default, awaiting input from the quantum segment only when necessary and then subsequently calculating a pressure correction. Such an approach not only exemplifies the efficient division of labor between quantum and classical computing resources but also underscores the potential for quantum-enhanced algorithms to integrate into and enhance existing computational workflows. 

\vspace{-\baselineskip}
\section*{Acknowledgements}

The authors would like to thank Dr Steven Lind at The University of Manchester for several insightful discussions. Similarly, the authors are grateful to Prof Vivien Kendon at University of Strathclyde for useful comments and discussion on this work, as well as the rest of the QEVEC/QuANDiE teams. The authors acknowledge funding through UKRI EPSRC projects (EP/W00772X/2 and EP/Y004515/1).
This work used the DiRAC@Durham facility managed by the Institute for Computational Cosmology on behalf of the STFC DiRAC HPC Facility (www.dirac.ac.uk). The equipment was funded by BEIS capital funding via STFC capital grants ST/K00042X/1, ST/P002293/1, ST/R002371/1 and ST/S002502/1, Durham University and STFC operations grant ST/R000832/1. DiRAC is part of the National e-Infrastructure.

\bibliographystyle{naturemag}
\bibliography{references} 
\clearpage
\onecolumn
\begin{center}
\vspace*{0.5cm}
    \LARGE{Supplementary Information For}\\
    \LARGE{Integrating Quantum Algorithms Into Classical Frameworks: A Predictor-corrector Approach Using HHL} \\[1em]
    \large Omer Rathore$^{1*}$, Alastair Basden$^{1}$, Nicholas Chancellor$^{1,2}$, Halim Kusumaatmaja$^{1,3}$ \\
    \normalsize
    $^{1}$ Department of Physics, Durham University, Durham DH1 3LE, United Kingdom \\
    $^{2}$ School of Computing, Newcastle University, Newcastle upon Tyne NE4 5TG, United Kingdom \\
    $^{3}$ Institute for Multiscale Thermofluids, School of Engineering, University of Edinburgh,\\
  Edinburgh EH9 3FD, United Kingdom \\ [2em]
\end{center}

A brief summary of the supplementary material included here is as follows: 
\begin{itemize}
    \item Supplementary A - Overview of HHL circuit
    \item Supplementary B - Details on how HHL output was simulated 
    \item Supplementary C - Mathematical background on incompressible smoothed particle hydrodynamics (ISPH)
    \item Supplementary D - Notes on asynchronous implementation of the proposed predictor-correctors
\end{itemize}
\section*{Supplementary A}
\begin{figure*}[h!]
\centering
\includegraphics[width=0.8\textwidth]{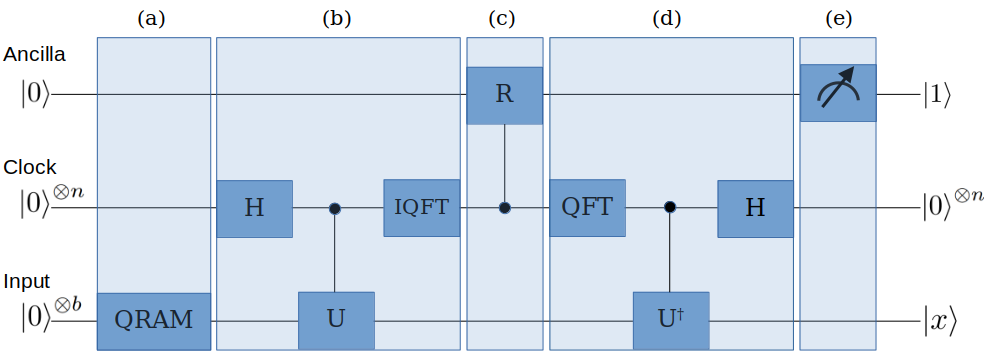}
\caption{Summary circuit diagram for the HHL algorithm. The five main components are (a) State preparation, (b) Quantum phase estimation, (c) Controlled rotation, (d) Inverse quantum phase estimation and (e) Ancilla qubit measurement. Quantum phase estimation consists of Hadamard gates (H), controlled unitaries (U) and the inverse quantum Fourier transform (IQFT). Meanwhile, the input consists of three registers, referred to here as the ancilla, clock and input registers.   }
\label{fig:1s}
\end{figure*}

\noindent The goal of the HHL algorithm is to find the solution to a Quantum Linear System Problem (QLSP), which is given by:

\begin{equation}\label{e:1s}
    \ket{x}=A^{-1}\ket{b}.
\end{equation}
By decomposing the matrix {A} into its eigenvalues $\lambda_i$ and eigenvectors $u_i$, as well as expressing $\Vec{b}$ in the basis formed by these eigenvectors, we can simplify Equation \ref{e:1s} to:

\begin{equation}\label{e:22s}
    \ket{x}=A^{-1}\ket{b}=\sum \lambda_i^{-1}b_i\ket{u_i}.
\end{equation}

The HHL algorithm achieves this solution using the quantum circuit illustrated in Figure \ref{fig:1s}. This circuit involves three input registers: the ancilla, clock, and input registers. Together, they initially form the state:

\begin{equation}
   \Psi =\ket{0}^{\otimes b}\ket{0}^{\otimes n}\ket{0}.
\end{equation}
The total number of qubits is composed of one ancilla qubit, $n$ clock qubits (which determine the precision for the phase estimation step), and the $b$ qubits required to encode the vector state $\Vec{b}$. The first step is to rotate the components of the input register in order to encode $\Vec{b}$ into the amplitudes. Assuming this can be achieved via QRAM, the resultant state is:

\begin{equation}
   \Psi_a =\ket{b}^{\otimes b}\ket{0}^{\otimes n}\ket{0}.
\end{equation}

Given a unitary operator with eigenstates $\ket {\zeta_i}$, and eigenvalues $e^{i \phi_i}$, QPE facilitates the mapping,
\begin{equation}
\ket{0}\ket{\zeta_i} \rightarrow \ket {\Tilde{\phi}_i}\ket{\zeta_i},
\end{equation}
where $\Tilde{\phi}_i$ is an approximation of the eigenvalue phase. This approximation is due to the binary nature of basis encoding, which limits precision based on the number of qubits in the clock register.

In HHL, the controlled unitary operation encodes the LSP matrix as its Hamiltonian ($U=e^{iAt}$), such that the eigenvalues of the gate are proportional to the eigenvalues of $A$. After QPE, the resultant state becomes:

\begin{equation}
    \Psi_b =\ket{b}^{\otimes b}\ket{\Tilde{\lambda}}^{\otimes n}\ket{0},
\end{equation}
where $\Tilde{\lambda}$ is an approximation to a scaled version of the eigenvalues of the LSP matrix. Subsequently a controlled rotation is applied to the ancilla qubit, conditioned on the clock register, resulting in the state,

\begin{equation}
    \Psi_c =\ket{b_i}^{\otimes b}\ket{\Tilde{\lambda_i}}^{\otimes n} \left(
    \sqrt{1-\frac{C^2}{\Tilde{\lambda}_i^2}}\ket{0} +\frac{C}{\Tilde{\lambda}_i}\ket{1}
    \right),
\end{equation}
with normalisation constant $C$.
To disentangle the input and clock registers, inverse QPE is applied, removing the now redundant information stored in the clock register. The state then becomes:

\begin{equation}
    \Psi_d =\ket{b_i}^{\otimes b} \left(
    \sqrt{1-\frac{C^2}{\Tilde{\lambda}_i^2}}\ket{0} +\frac{C}{\Tilde{\lambda}_i}\ket{1}
    \right).
\end{equation}

Expressing $\ket{b}$ in the eigenbasis of $A$ (i.e. $\ket{b}=\sum b_i \ket{u_i}$, it is evident by only selecting the state corresponding to an ancilla bit measurement of 1, the final state is 

\begin{equation}
        \Psi_e =b_i\ket{u_i} \frac{C}{\Tilde{\lambda}_i}\ket{1},
\end{equation}
 which corresponds to the desired solution as defined in Equation \ref{e:22s}. For a more detailed walk-through the interested reader is refereed to relevant literature \cite{zaman2023step,dervovic2018quantum}.

\section*{Supplementary B}

In this work, the predictor-corrector methods hinge on the ability to reproduce outputs from HHL. Unfortunately, the limitations of current quantum hardware preclude direct quantum simulation of the HHL circuit. Nevertheless, since our methods focus primarily on processing the output of HHL, simulating these outputs directly is a viable and justifiable approach.

This simulation is done here by recognising that the completion of the HHL circuit results in a quantum state characterized by a superposition of probability amplitudes, each encoding components of the solution. By solving the system classically \emph{a priori}, we can construct a probability distribution whose weights are directly proportional to these probability amplitudes squared. We then perform random sampling from this pre-computed distribution to simulate the outcomes of individual measurements on the HHL solution vector. 

For example, consider the simple case of two variables with an output state from HHL being described by,  

\begin{equation}
    \Psi=a_1\ket{0}+a_2\ket{1}.
\end{equation}
The probability amplitudes are in the same ratios as the solution of the LSP ($x_i$), 

\begin{equation}
    \frac{a_1}{a_2}=\frac{x_1}{x_2}.
\end{equation}
Thus, we first calculate the solution components, $x_i$, 
using classical methods. This enables us to establish a probability distribution weighted according to $x_i^2$
. Sampling from this distribution effectively simulates measuring the quantum state, which probabilistically collapses to a basis state with probabilities  $a_i^2$.

One challenge with this approach is the inability to distinguish between positive and negative probability amplitudes since we are sampling from a distribution corresponding to probability amplitudes \emph{squared}. This is addressed here by decomposing the initial LSP into two separate component LSP systems, each representing different scenarios of amplitude sign combinations. These are given by 

\begin{equation}
    A\Vec{x}_1=\Vec{b}_1, \quad A\Vec{x}_1=\Vec{b}_2,
\end{equation}
where $\Vec{b}_1$ and $\Vec{b}_2$ contain only the negative and positive components of the original $\Vec{b}$ vector as non-zero values. This ensures that components of $\Vec{x}_1$ and $\Vec{x}_2$ are always positive.
\section*{Supplementary C}

Incompressible SPH (ISPH) is a particle based method that evolves the Navier-Stokes in Lagrangian form. To enforce incompressibility, a projection method \cite{cummins1999sph} is commonly employed in order to ensure a divergence free velocity field. The algorithm first advects particles to an intermediate position ($\mathbf{x}^*$) using the velocity at the current time step ($\mathbf{u}^n$),
\begin{equation}\label{z:1}
    \mathbf{x}_i^*=\mathbf{x}_i^n+\mathbf{u}_i^n\Delta t.
\end{equation}
An intermediate velocity is then calculated at this position using the contribution of viscous forces, 
\begin{equation}
\mathbf{u}_i^*=\mathbf{u}_i^n+\nu\nabla^2\mathbf{u}_i^n\Delta t .
\end{equation}
The pressure at the next time step is obtained by solving an implicit pressure-Poisson equation, 
\begin{equation}\label{e:3s}
    \nabla \cdot \left( \frac{1}{\rho}\nabla P^{n+1}\right)_i=\frac{1}{\Delta t}\nabla\cdot\mathbf{u}_i^*.
\end{equation}
The velocity is then computed by projecting the intermediate velocity onto a divergence free space to ensure incomprehensibility,
\begin{equation}\label{e:4}
    \mathbf{u}_i^{n+1}=\mathbf{u}_i^*-\left(  \frac{1}{\rho}\nabla P_i^{n+1} + g \right) \Delta t.
\end{equation}
Lastly, the particles are advected to their positions for the subsequent time step, 
\begin{equation}\label{z:2}
    \mathbf{x}_i^{n+1}=\mathbf{x}_i^n+\left( \frac{\mathbf{u}_i^{n+1} +\mathbf{u}_i^n}{2}\right)\Delta t .
\end{equation}

Of particular importance to this work is Equation \ref{e:3s}, as when discretised it results in the LSP used as a target for our predictor-corrector. Central to discretisation in SPH is the integral approximation to a function over domain $\Omega$ at location $x$, as given by 
\begin{equation}\label{ee:1}
    f(x) \approx \int_\Omega f(x')W(x-x',h)dx',
\end{equation}
where $W$ is the weighting (i.e. kernel) function within a zone of influence around the point $x$, that is determined by the smoothing length $h$. The value of a function at a point is thus given by an interpolation over neighbouring particles, weighted by the kernel function. Kernels are designed to have compact support as controlled by the smoothing length. 

The discrete approximation of Equation \ref{ee:1} is the standard SPH summation of the form
\begin{equation}
   \langle f(x)\rangle = \sum^N_j f(x_j)W(x-x_j,h)V_j,
\end{equation}
where $\langle \rangle$ indicates an interpolated value. This can be further simplified to 
\begin{equation}
    f(x_i)=\sum _j^N \frac{m_j}{\rho_j}f_jW_{ij},
\end{equation}
where the volume $V_j$ is expressed in terms of mass and density, $W_{ij}=W(x_i-x_j,h)$, $f_j=f(x_j)$ and the $\langle \rangle$ has been dropped for convenience. Using this summation it is possible to arrive at the following expressions for gradient and Laplacian operators \cite{lind2020review,morris1997modeling}, 
\begin{equation}
    \nabla \phi_i=-\sum_j^N V_j(\phi_i-\phi_j)\nabla_iW_{ij},
\end{equation}
\begin{equation}
(\nabla\cdot a \nabla \mathbf{b})_i=\sum_j^N\frac{m_j(a_i+a_j)x_{ij}\cdot\nabla_iW_{ij}}{\rho_j(r_{ij}^2+\eta^2)}\mathbf{b}_{ij},
\end{equation}
where $\eta$ is a small parameter used to avoid singularities. Substituting this expression for the laplacian into Equation \ref{e:3s}
results in a LSP of the form $\mathbf{A}\Vec{P}=\Vec{b}$, where the known vector $\Vec{b}$ contains the divergence of the velocity field and the matrix $\mathbf{A}$ represents particle-particle interactions. Each row of this matrix correponds to a particular particle and each column for that row represents an interaction with a neighbour. As the particles keep on moving and their zone of influence changes, this matrix needs to be updated and solved again to ensure the correct pressure is being used for the projection in Equation \ref{e:4}. 
\section*{Supplementary D}

\begin{figure*}[h!]
\centering
\includegraphics[width=0.8\textwidth]{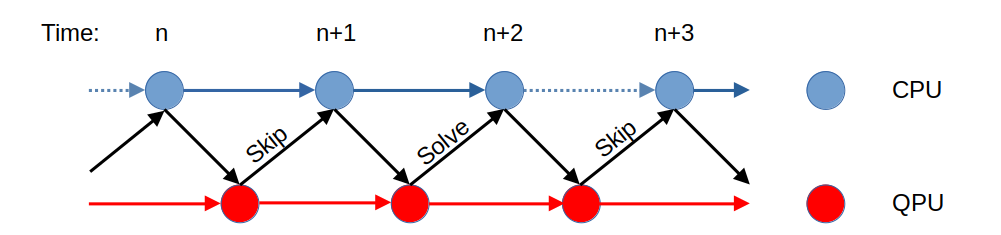}
\caption{Illustration of asynchronous predictor-corrector framework. The CPU (blue) proceeds with the ISPH algorithm without updating pressure by default. At each time step, a QPU (red) asynchronously evaluates whether a pressure correction is needed or not and feeds forward the decision for the next time step to the CPU stack. This can then result in a subsequent recalculation of the pressure if needed, as indicated by the dotted arrows. As such, there is no need for the CPUs to wait on a decision from the \emph{predictor} component of our predictor-corrector.  }
\label{fig:2}
\end{figure*}

\noindent An advantage of our proposed predictor-corrector approach lies in its ability to operate asynchronously across heterogeneous hardware configurations. This effectiveness stems from utilizing the quantum processing unit (QPU) as a specialized tool designated solely for specific tasks, while the main body of the algorithm continues to execute on classical processing units (CPUs). In practice, this could involve resolving Equations \ref{z:1}-\ref{z:2} on the CPUs, without concurrently updating the pressure per Equation \ref{e:3s}. Consequently, the classical portion of the algorithm operates under the assumption that a pressure update is not immediately necessary. However, at each time step, the necessary inputs are dispatched to the QPU, which then determines whether a pressure update is required. This decision is made in coordination with the CPUs, and the results are fed back at a subsequent time step. This division of labor between QPUs and CPUs, as illustrated in Figure \ref{fig:2}, could potentially enhance the system's efficiency by leveraging the unique capabilities of each type of processor.

\begin{figure*}[h!]
\centering
\includegraphics[width=0.6\textwidth]{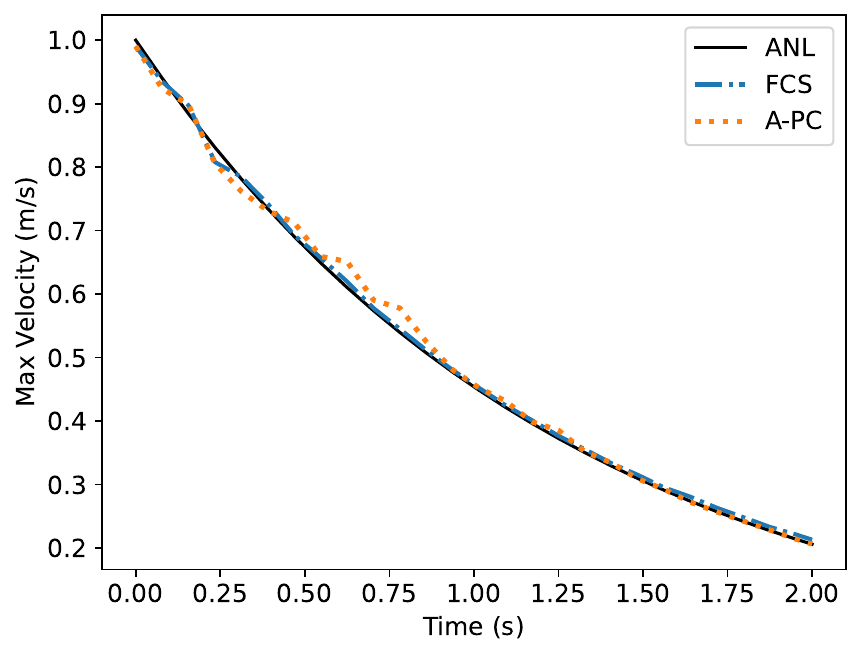}
\caption{Decay profiles for maximum velocity in the domain for the Taylor-Green Vortex. Results shown include those for the analytical (ANL), full classical solve (FCS) i.e. solving for pressure at every time step and asynchronous predictor-corrector (A-PC). }
\label{fig:3}
\end{figure*}

As a preliminary demonstration, this asynchronous implementation was emulated using the Taylor-Green Vortex (TGV), where the decision regarding whether to skip or evaluate the LSP was made based on the known information at the previous time step. For example, the decision to skip at time $t_n$ was made at $t_{n-1}$ using the predicted distribution at $t_{n-1}$ and the last known solution (e.g. at $t_{n-2}$). As a result of this, the CPU stack is aware whether to skip the LSP or not at the start of step $t_n$ and does not need to wait for a decision. Figure \ref{fig:3} illustrates how there is still very good agreement with the predicted decay profiles. There are some small oscillations following the initial SPH kink as there is delay between determining the need for a pressure update and actually evaluating this update. However, this quickly self-corrects, resulting in excellent overall agreement. A more detailed investigation of parameter requirements or the effects of further staggering the process is left for future works.    

\end{document}